%% file: Deschle_etal_InformationInChimeras_Revision.tex
\documentclass[pre,twocolumn,superscriptaddress]{revtex4-1}

\usepackage[utf8x]{inputenc}

\usepackage{amssymb}
\usepackage{amsmath}
\DeclareMathOperator*{\argmax}{\arg\max}
\usepackage[hyperfootnotes=true,colorlinks]{hyperref}
    	\hypersetup{colorlinks,%
    	citecolor=blue,%
    	filecolor=green,%
    	linkcolor=blue,%
    	urlcolor=blue,%
   	}

\usepackage{graphicx}
\usepackage{mathtools}
\usepackage[tables]{xcolor}

\newcommand{\taumax}{\tau_{\rm max}}
\newcommand{\Imax}{I_{\rm max}}
\renewcommand{\SS}{{\rm{SS}}}
\newcommand{\SD}{{\rm{SD}}}
\newcommand{\DS}{{\rm{DS}}}
\newcommand{\ISSmax}{\Imax^\SS}
\newcommand{\ISDmax}{\Imax^\SD}
\newcommand{\avt}[1]{\langle #1\rangle_t}
\newcommand{\avo}[1]{\langle #1\rangle_j}

\begin{document}
\title{Directed flow of information in chimera states}

\author{Nicol\'as Deschle}%
\email{n.deschle@vu.nl}
\affiliation{Faculty of Behavioural and Movement Sciences, Amsterdam Movement Sciences \& Institute for Brain and Behavior Amsterdam, Vrije Universiteit Amsterdam, van der Boechorststraat 9, Amsterdam 1081 BT, The Netherlands}
\affiliation{Institute for Complex Systems and Mathematical Biology, University of Aberdeen, King's College, Old Aberdeen AB24 3UE, United Kingdom}

\author{Andreas Daffertshofer}
\email{a.daffertshofer@vu.nl}
\affiliation{Faculty of Behavioural and Movement Sciences, Amsterdam Movement Sciences \& Institute for Brain and Behavior Amsterdam, Vrije Universiteit Amsterdam, van der Boechorststraat 9, Amsterdam 1081 BT, The Netherlands}

\author{Demian Battaglia}
\email{demian.battaglia@univ-amu.fr}
\affiliation{Institute for Systems Neuroscience, University Aix-Marseille, Boulevard Jean Moulin 27, 13005 Marseille, France}
\author{Erik A. Martens}%
\email{eama@dtu.dk}
\affiliation{Department of Applied Mathematics and Computer Science, Technical University of Denmark, 2800 Kgs. Lyngby, Denmark}
\affiliation{Department of Biomedical Sciences, University of Copenhagen, Blegdamsvej 3, 2200 Copenhagen, Denmark}
\date{\today}

\begin{abstract}
We investigated interactions within chimera states in a phase oscillator network with two coupled subpopulations. 
To quantify interactions within and between these subpopulations, we estimated the corresponding (delayed) mutual information that -- in general -- quantifies the capacity or the maximum rate at which information can be transferred to recover a sender's information at the receiver with a vanishingly low error probability.
After verifying their equivalence with estimates based on the continuous phase data, we determined the mutual information using the time points at which the individual phases passed through their respective Poincar\'e sections. This stroboscopic view on the dynamics may resemble, e.g., neural spike times, that are common observables in the study of neuronal information transfer. This discretization also increased processing speed significantly, rendering it particularly suitable for a fine-grained analysis of the effects of experimental and model parameters. In our model, the delayed mutual information within each subpopulation peaked at zero delay, whereas between the subpopulations it was always maximal at non-zero delay, irrespective of parameter choices.
We observed that the delayed mutual information of the desynchronized subpopulation preceded the synchronized subpopulation. Put differently, the oscillators of the desynchronized subpopulation were 'driving' the ones in the synchronized subpopulation. These findings were also observed when estimating mutual information of the full phase trajectories. We can thus conclude that the delayed mutual information of discrete time points allows for inferring a \emph{functional} directed flow of information between subpopulations of coupled phase oscillators.
\begin{description}
\item[Keywords]
Chimeras, phase oscillators, coupled networks, mutual information, information flow
\end{description}
\end{abstract}

\pacs{Valid PACS appear here}

\maketitle

\section{Introduction}
Oscillatory units are found in a spectacular variety of systems in nature and technology. Examples in biology include flashing fireflies \cite{ermentrout1984beyond}, cardiac pacemaker cells \cite{guevara1981phase,guevara1989alternans,glass1988clocks,zeng1990theoretical,glass1991low}, and neurons \cite{singer1999neurobiology,gray1989oscillatory,mackay1997synchronized,singer1995visual, stern1998membrane}; in physics one may think of Josephson junctions \cite{jain1984mutual, saitoh1991phase, valkering2000dynamics}, electric power grids \cite{menck2014dead,nardelli2014models,motter2013spontaneous,dorfler2005synchronization,lozano2012role,susuki2011nonlinear,rohden2012self}, and, of course, pendulum clocks \cite{huygens1673horologium}. 
Synchronization plays an important role in the collective behavior of and the communication between individual units~\cite{pikovsky2003synchronization,buzsaki2006rhythms,Strogatz2003}. 
In the last two decades or so, many studies addressed the problem of synchronization in networks with complex structure, such as networks of networks, hierarchical networks and multilayer networks~\cite{boccaletti2014structure, arenas2008synchronization,kivela2014multilayer,rodrigues2016kuramoto}. 
Alongside efforts studying synchronization on networks, a new symmetry breaking regime coined \emph{chimera state} has been observed. In a chimera state an oscillator population `splits' into two parts, one being synchronized and the other being desynchronized~\cite{kuramoto2002coexistence, abrams2004chimera, montbrio2004synchronization}, or more generally,  different levels of synchronization~\cite{MartensPanaggioBick2016}. This state is a striking manifestation of symmetry breaking, as it may occur even if oscillators are identical and coupled symmetrically; see~\cite{panaggio2016chimerastates,Scholl2016} for recent reviews. Chimera states have spurred much interest resulting in many theoretical investigations, but they have also been demonstrated in experimental settings using, e.g., mechanical and (electro-)chemical oscillators or  lasers~\cite{Martens2013,Tinsley2012,Hagerstrom2012,Wickramasinghe2013}, and electronic circuits implementing FitzHugh-Nagumo neurons~\cite{Gambuzza2014}. 

Chimera states can be considered  patterns of localized synchronization. As such they may contribute to the coding of information in a network. This is particularly interesting since systems with chimera states typically display multi-stability, i.e., different chimera configurations may co-exist for identical parameters~\cite{martens2010bistable,Martens2010var,shanahan2010metastable,Bick2017c}. Such a network may hence be able to encode different stimuli through different chimera states without the need for adjusting parameters or altering network structure. It is even possible to dynamically switch between different synchronization patterns, thus allowing for dynamic coding of information~\cite{BickMartens2015,martens2016basins}.

The mere existence of different levels of synchronization in the same network can also facilitate the transfer of information across a network, especially between subpopulations. Coherent oscillations between neurons or neural populations have been hypothesized to provide a communication mechanism while full neural synchronization is usually considered pathological~\cite{buzsaki2006rhythms,fries2005a,fell2011role}.  A recent study reported chimera-like states in neuronal networks in humans, more specifically in electro-encephalographic patterns during epileptic seizures~\cite{andrzejak2016all}.
Admittedly, our understanding of these mechanism is still in its infancy also because the (interpretations of) experimental studies often lack mathematical rigor, both regarding the description of synchronization processes and the resulting implications for the network's capacity for information processing.
\emph{What is the information transfer within and between synchronized and desynchronized populations?}
We investigated the communication channels between two subpopulations in a system of coupled phase oscillators. Since we designed that system to exhibit chimera states, we expected non-trivial interactions between the subpopulations. To tackle this, we employed the delayed version of mutual information which measures the rate at which information can be sent and recovered with vanishingly low probability of error \cite{cover2012elements}. Furthermore, assuming symmetry of our configuration, any directionality found in the system should be regarded of functional nature. With this we sought not only to answer the aforementioned question but to contribute to a more general understanding of how information can be transferred between subpopulations of oscillators. 
In view of potential applicability in other (experimental) studies, we also investigated whether we could gain sufficient insight into this communication by looking at the passage times through the oscillators' Poincar\'e sections rather than evaluating the corresponding  continuous-time data. Such data may resemble, e.g., spike trains in neurophysiological assessments. 

Our paper is structured as follows. First, we introduce our model in the study of chimera states~\cite{Abrams2008,montbrio2004synchronization,panaggio2016chimera,Bick2018}. It consists of two subpopulations that either can be synchronized or desynchronized. We generalize this model by including distributed coupling strengths among the oscillatory units as well as additive Gaussian white noise. We briefly sketch the conditions under which chimera states can exist. 
Second, we outline the concept of \emph{delayed} mutual information and detail how to estimate this for `event-based' data where we define events via Poincar\'e sections of the phase oscillators' trajectories.
With this tool at hand, we finally investigate the flow of information within and between the two subpopulations, and characterize its dependency on the essential model parameters.

\section{Model}\label{sec:model}
We build on a variant of the noisy Kuramoto-Sakaguchi model~\cite{Acebron2005}, generalized to $M$ subpopulations of phase oscillators~\cite{BickLaingGoodfellowMartens2019}. The phase $\phi_{j,\mu}(t)$ of oscillator $j=1,\ldots,N$ in population $\mu=1,\ldots,M$ evolves according to
\begin{widetext}
\begin{align}\label{model}
d\phi_{j,\mu}(t) = \left[
\omega_{j,\mu} +
\sum_{\nu=1}^M\sum_{k=1}^{N_\mu}C_{kj,\mu\nu}\sin\left(\phi_{k,\nu}(t)-\phi_{j,\mu}(t) - \alpha\right)
\right]dt
+dW_{j,\mu}(t) \ ,
\end{align}
\end{widetext}
where $\omega_{j,\mu}$ is the natural frequency of the oscillator $j$ in subpopulation $\mu$. Throughout our study $\omega_{j,\mu}$ is drawn from a zero-centered Gaussian distribution with variance $\sigma_{\omega}^2$. 
Oscillator $j$ in population $\mu$ and oscillator $k$ in population $\nu$ interact sinusoidally with coupling strength $C_{kj,\mu\nu}$ and a phase lag $\alpha$. The phase lag $\alpha$ varies the interaction function between more cosine ($\alpha\to\pi/2$) or more sine like-behavior ($\alpha\to0$) and can be interpreted as a (small) transmission delay between units~\cite{panaggio2016chimerastates}.
The additive noise term $dW_{j,\mu}(t)$ represents mean-centered Gaussian noise with variance (strength) $\sigma^2_W$, i.e., $\avt{dW_{j,\mu}(t)dW_{k,\nu}(t')} = \sigma^2_W\delta_{jk}\delta_{\mu\nu}\delta(t-t')$.

For the sake of legibility, we restrict this model to the case of $M=2$ subpopulations of equal size $N_1=N_2=N$. In the absence of noise and in the case of identical oscillators, uniform phase lag and homogeneous coupling, i.e., for $\sigma_W=0$,  $\sigma_\omega=0$, $\alpha_{\mu\nu} \coloneqq \alpha$, and $\sigma_C=0$, respectively, the aforementioned studies establishes corresponding bifurcation diagrams \cite{Abrams2008,montbrio2004synchronization,panaggio2016chimera}.

To generalize the model towards more realistic, experimental setting, we included distributed coupling strengths via a non-uniform, (statistically) symmetric coupling between subpopulations. This can be cast in the following coupling matrix:
\begin{align}
  C= \frac{1}{N}
  \begin{pmatrix}
    C_{\zeta} & C_{\eta} \\
    C_{\eta} & C_{\zeta}
  \end{pmatrix} \in \mathbb{R}^{2N\times2N}
\end{align}
with block matrices $C_{\zeta},C_{\eta} \in R^{N \times N}$. The self-coupling within the subpopulation and the neighbor coupling between subpopulations are represented by $C_\zeta$ and $C_\eta$, respectively. Each block $C_{\zeta}$ (or $C_{\eta}$) is composed of random numbers drawn from a normal distribution with mean $\zeta$ (or $\eta$) and variance $\sigma_C^2$. By this, we can tune the heterogeneity in the network. 
We would like to note that, in general, the chosen coupling is only symmetric for $\sigma_C>0$ in the sense of a statistical average, and that the expected coupling values $\zeta$ and $\eta$ in each block only can be retrieved by the mean in the limit of large numbers. Although the symmetry between the two subpopulations $ 1\leftrightarrow 2$ is broken and only preserved in a statistical sense in the limit of large numbers, we verified that chimera states appeared in both configurations. That is, subpopulation~1 is synchronized and subpopulation~2 is desynchronized ({\it\SD}) and subpopulation 2 is synchronized and subpopulation 1 is desynchronized ({\it\DS}), for a particular choice of parameters using numerical simulations. 
Another consequence of distributed coupling strengths is that only a very few oscillators may have experienced negative (inhibitory) coupling, which can be neglected.

Following~\cite{Abrams2008}, we parametrize the relation between strengths by $A = \zeta - \eta$ with $\zeta + \eta = 1$ and the phase lag $\alpha_{\mu,\nu}$ by $\beta_{\mu,\nu} = \frac{\pi}{2}-\alpha_{\mu,\nu}$ which here can be kept homogeneous for the entire network, i.e., $\beta_{\mu,\nu}  \coloneqq  \beta$. By this, we may recover the results reported  in~\cite{Abrams2008,montbrio2004synchronization} in the limit $\sigma_C\to0$, under the proviso of $ \sigma_W = \sigma_\omega = 0$. 
In brief, increasing $A$ from $0$ while fixing $\alpha$ (or $\beta \coloneqq \pi/2-\alpha$) yields the following bifurcation scheme: a ``stable chimera'' is born through a saddle-node bifurcation and becomes unstable in a supercritical Hopf bifurcation with a stable limit cycle corresponding to a ``breathing chimera'', which eventually is destroyed in a homoclinic bifurcation. For all parameter values, the fully synchronized state exists and is a stable attractor; see also Fig.~\ref{fig:App1} in  {\it{Appendix}}~\ref{app:stabilitydiagram}. 
Subsequent studies demonstrated robustness of chimera states against non-uniformity of delays ($\alpha_{\mu \nu}\neq\alpha$)~\cite{MartensPanaggioBick2016}, heterogeneity of oscillator frequencies ($\sigma_{\omega}>0$)~\cite{Laing2009b}, and additive noise~\cite{laing2012disorder}. 
Although non-uniform phase lags lead to less degenerate dynamics and give room for more complex dynamics~\cite{MartensPanaggioBick2016,Bick2018}, we restrict ourselves to the case of uniform phase lags without compromising the essential phenomenology of chimera states.

\begin{figure}[!h]
\begin{center}
    \includegraphics[width = \columnwidth]{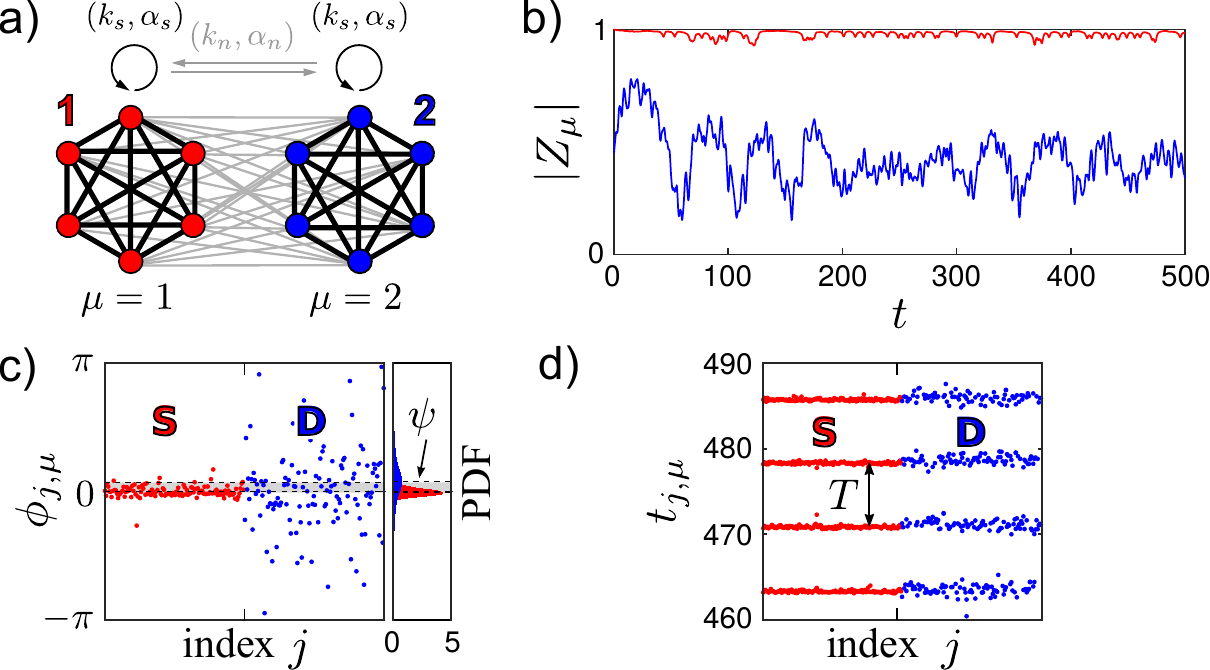}
  \caption{\label{fig:chimerastate}
  Panel (a): Chimera state in a network with $ M = 2 $ oscillator subpopulations and non-uniform coupling, simulated using Eq.~\eqref{model}. 
  Panel (b): The time evolution of the order parameter, $|Z_\mu(t)|, \mu=1,2$, reflects the fluctuations driven by the coupling and the external field.
  Despite temporary deviations off the synchronization manifold $ |Z_1| = 1 $, the system remains near the chimera state attractor. 
  That is, chimera states appear robust to both coupling heterogeneity and additive noise. Considering the time-average of the order parameters, the state may be classified as a stable chimera.
  %
  %
  Panel (d): Times $t'_{j,\mu}$ correspond to the Poincar\'e sections of the phases i.e., $t'_{j,\mu}$ such that $\phi_{i}(t'_{j,\mu})=0$. 
  The time period of collective oscillation is defined as $T \coloneqq 1/\Omega$ and $\Omega \coloneqq \avo{\avt{\frac{d \phi}{dt}} }$ where $\avt{\cdot}$ and $\avo{\cdot}$ denote averages over time and oscillators, respectively. 
  Parameters are: $ A=0.275 $, $ \beta=0.125 $, $ \sigma_C = 0.01$, $ \sigma_\omega = 0.01 $, $\sigma_W^2 = 0.001$.
  }
\end{center}
\end{figure}

The macroscopic behavior, i.e. the synchronization of the two populations, may be characterized by complex-valued order parameters, which are either defined on the population level,  i.e., $Z_\mu \coloneqq N^{-1}\sum_{j=1}^N e^{i\theta_{\mu,j}}$, or globally, i.e., $Z=(2N)^{-1}\sum_{\mu=1}^M\sum_{j=1}^N e^{i\theta_{\mu,j}}$. As common in the studies of coupled phase oscillators, the level of synchrony can be given by $R_\mu \coloneqq |Z_\mu|$. Thus, $R_\mu=1$ implies that oscillators in population $\mu$ are perfectly phase synchronized ({\rm{S}}), while for $R_\mu<1$ the oscillators are imperfectly synchronized or de-synchronized ({\rm{D}}). For our two-subpopulation case, full synchrony (\SS) hence occurs when $R_1\lessapprox 1$ and $R_2\lessapprox 1$, and chimera states are present if $R_1<1$ and $R_2\lessapprox 1$ or vice versa. The angular order parameter, $\Phi_\mu \coloneqq \arg{{Z}_\mu}$ keeps track of the average phase of the (sub)population. Fluctuations inherent to the model may affect the order parameter as illustrated in Fig.~\ref{fig:chimerastate}(b) (please refer to section \ref{simulations} for the numerical specifications). We therefore always considered averages over time, ${\avt{R_\mu}}$, when discussing the stability of a state.
In fact, in our model chimera states remain stable for relatively large coupling heterogeneity, $\sigma_C>0$ presuming $\sigma_\omega > 0$ and $\sigma_W >0$, as is evidenced by numerical simulations; see also {\it{Appendix}}~\ref{app:stabilitydiagram}, Fig.~\ref{fig:App1}. The perfect synchronization manifold with $R=1$ cannot be achieved; see Fig.~\ref{fig:chimerastate}(b). Further aspects of these noisy dynamics will be presented elsewhere.

Adding noise and heterogeneity to the system may alter its dynamics. In the present work we concentrated on parameter regions characterized by the occurrence of dynamic states that did resemble stable chimeras, i.e., $\avt{R_1} > \avt{R_2}$, where $\avt{ \cdot} $ denotes the average over a duration of $T = 9 \cdot 10^6$ time steps (after removing a transient of $T = 10^5$ time steps). Fig.~\ref{fig:App1} in {\it{Appendix}}~\ref{app:stabilitydiagram} provides an overview and explanation of how parameter points $A$ and $\beta$ were selected. 
 
\section{Implementation and analysis}

\subsection{Simulations} \label{simulations}
For the numerical implementation we employed a Euler-Maruyama scheme with $\Delta{t}=10^{-2}$ for $N=128$ phase oscillators per subpopulation that were evolved for $T=10^6$~\cite{Acebron2005}. We varied the coupling parameter $A$ and the phase lag parameter $\beta$, while we fixed the width (standard deviation) of the natural frequency distribution to $ \sigma_{\omega} = 0.01 $ and of the coupling distribution to $ \sigma_C = 0.01 $. The additive noise had variance $\sigma_W^2 = 0.001$. 

\subsection{Mutual Information}
Mutual information $I(X; Y)$, first introduced by Shannon~\cite{Shannon1949}, is meant to assess the dependence between two random variables $X$ and $Y$. It measures the amount of information that $X$ contains about $Y$. In terms of communication theory, $Y$ can be regarded as output of a \emph{communication channel} which depends probabilistically on its input, $X$. By construction, $I(X; Y)$ is non-negative; it is equal to zero if and only if $X$ and $Y$ are independent. Moreover, $I(X; Y)$ is symmetric, i.e., $I(X;Y)=I(Y;X)$ implying that it is not directional. The mutual information may also be considered as a measure of the reduction in the uncertainty of $X$(Y) due to the knowledge of $Y$(X) or, in terms of communication, the rate at which information is being shared between the two~\cite{cover2012elements}. The maximum rate at which one can send information over the channel and recover it at the output with a vanishingly low probability of error is called \emph{channel capacity}, $c = \max_{p(x)} I (X; Y)$.
In networks with oscillatory nodes, the random variables are the degree of synchronization of different subpopulations, and the rate of information shared will depend on the state of the system for each of the subpopulations. 

Mutual information can be defined via the Kullback-Leibler divergence or in terms of entropies,
\begin{align} \nonumber
  I(X;Y) 
  &= H(X)-H(X|Y) \\ &= H(X) + H(Y) - H(X,Y).
\end{align}
where $H(X)$ is the entropy of the random variable $X$, i.e., $ H(X)=-\int dx\ p_X(x)\log p_X(x)$ with $p_X(x)$ being the corresponding probability density. 

\subsection{Delayed mutual information \label{sec:methods:delayedMI}}
For time-dependent random variables, one may generalize this definition to that of a \emph{delayed mutual information}. This variant dates back to Fraser and Swinney \cite{FraserSwinney1986}, who used the delayed mutual information for estimating the embedding delay in chaotic systems | in that context the delayed \emph{auto}-mutual information is often considered a `generalization' of the auto-correlation function~\cite{abarbanel2012analysis}. Applications range from the study of coupled map lattices~\cite{kaneko1986lyapunov}  via spatiotemporal~\cite{VastanoSwinney1988} and general dependencies between time series~\cite{green1991dependent} to the detection of synchronization~\cite{palus1997detecting}. 
With the notation $I_{XY} \coloneqq I(X;Y)$ and $H_{X} \coloneqq H(X)$, the delayed mutual information  for a delay $\tau$ may read
\begin{align} \nonumber
I_{XY}(\tau) &= I(X(t);Y(t+\tau)) \\ &=H_X + H_Y - H_{XY}(\tau)
\label{eq:delayedMutualInformation}
\end{align}
which has the symmetry $I_{XY}(\tau) = I_{YX}(-\tau)$~\footnote{This follows because of:
  $\tilde{I}_{XY}(\tau) 
  =I\left(X(t);Y(t+\nu)\right)=\tilde{I}\left(X(t-\nu);Y(t)\right)
  =\tilde{I}\left(Y(t);X(t-\nu)\right)
  =\tilde{I}_{YX}(-\tau)$.}.
With this definition, one can measure the rate of information shared between $X$ and $Y$ as a function of time delay $\tau$. In fact, we are not particularly interested in the specific value of the mutual information but rather focus here on the time delay at which the mutual information is maximal. Hence, we define $\taumax \coloneqq \argmax_\tau I_{XY}(\tau)$. A positive (negative) value of $\taumax$ implies that $Y$ shares more information with a delayed (advanced) $X$. This means there is an information flow from $X$(Y) to $Y$(X).

\subsection{Delayed mutual information between subpopulations}
\paragraph{Estimates using continuous-time data.} When the time-dependent random variables are continuous time series $u_{\mu}(t)$ and $v_{\nu}(t)$ associated to populations $\mu$ and $\nu$, the delayed mutual information can be estimated from Eq.~\eqref{eq:delayedMutualInformation} for $X(t)=u_{\mu}(t)$ and $Y(t)=u_{\nu}(t)$,
\begin{align}\label{eq:MIcontinous}
  \tilde{I}_{\mu \nu}(\tau)=  \tilde{I}\left(u_{\mu}(t);u_{\nu}(t+\tau)\right) .
\end{align}
We determined the probability densities using kernel density estimators with Epanetchnikov kernels with bandwidths given through a uniform maximum likelihood cross-validation search. For our parameter settings (254 oscillators and $10^6$ samples), the resulting bandwidths ranged from about $0.10\ \textrm{rad}$ to $0.18 \ \textrm{rad}$. This software implementation is part of the \href{https://www.ics.uci.edu/~ihler/code/kde.html}{KDE-toolbox}; cf. \cite{ihler2003kernel} and \cite{thomas2014efficient} for alternative schemes.

\paragraph{Estimates using event-based time data.} \label{sec:discrete}
For the aforementioned event signals in the subpopulations $1$ and $2$, i.e., discrete time points are defined as passing moments through the respective Poincar\'e sections. The probability densities to incorporate when estimating the mutual information are densities of events, or densities of times. We implemented the probability estimates as follows. Let $S_\mu$ be a set of event times, i.e.,  $S_\mu=\{t^{(m)}_{\mu,1},\dots, t^{(m')}_{\mu,N_\mu}\}$ where $t^{(m)}_{\mu,i}$ stands for the time of the $m$-th event of oscillator $i$ in subpopulation $\mu$. Then, the probability density for an event to happen at time $t$ in subpopulation $\mu$ is $p_{\mu}(t) = P(t\in S_{\mu})$
 and the probability of an event to happen at time $t$ in subpopulation $\mu$ and time $t+\tau$ in subpopulation $\nu$ is 
 \begin{align}\label{eq:jointP} \nonumber
	p_{\mu\nu}(t,t+\tau) &= P\left(\left\{t\in S_{\mu}\right\} \cap \left\{t+\tau\in S_{\nu}\right\}\right)
				\\&= P\left(t\in S_\mu\right) + P\left(t+\tau\in S_\nu\right) \nonumber
				\\& \hspace{30pt} - P\left(\left\{t\in S_\mu\right\} \cup \left\{t+\tau\in S_\nu\right\}\right).
\end{align}
The delayed mutual information can be given as
\begin{align} \nonumber
  I_{\mu \nu}(\tau)
  &= \int dt \ p_{\mu \nu}(t,t+\tau)\log\dfrac{p_{\mu \nu}(t,t+\tau)}{p_{\mu}(t)p_{ \nu}(t)} \\ &= H_{\mu} + H_{\nu} - H_{\mu \nu}(\tau) \ .\label{eq:MIevent}
\end{align}
We again computed the probability densities using kernel density estimators \cite{ihler2003kernel} but now involving Gaussian kernels. We also adjusted the bandwidth selection to a spherical, local maximum likelihood cross-validation due to the sparsity of the data and the resulting bandwidths ranged from about $25$ to $35$ time units. These results appeared robust when using the aforementioned uniform search; again we employed the \href{https://www.ics.uci.edu/~ihler/code/kde.html}{KDE-toolbox}.

\subsection{Events defined via Poincar\'e sections}
\label{sec:methods:poincare}
We analyzed the times $t_{1,\mu}, t_{2,\mu}, \dots t_{N,\mu}$ at which the individual phases $\phi(t)_{j,\mu}$ passed through their respective Poincar\'e sections. The latter were defined as $t_{j,\mu}\in{\mathbb{R}}^+_0: \phi_{j,\mu}(t_{j,\mu})/2\pi \in {\mathbb{Z}}$. As mentioned above, every subpopulation $\mu$ generated an `event sequence' $S_\mu=\{t^{(m)}_{\mu,1},\dots, t^{(m')}_{\mu,N_\mu}\}$; which, as already said, may be considered reminiscent of spike trains; cf. Fig.~\ref{fig:chimerastate}(c).

\section{Results\label{sec:results}}

To determine the directionality of the information flow in the network we computed the time lagged mutual information within the subpopulations, $ I_{11}(\tau) $ and $ I_{22}(\tau) $ and between them, $ I_{12}(\tau) = I_{21}(-\tau) $.

\begin{figure}[!htp]
  \begin{center}
    \includegraphics[width = \columnwidth]{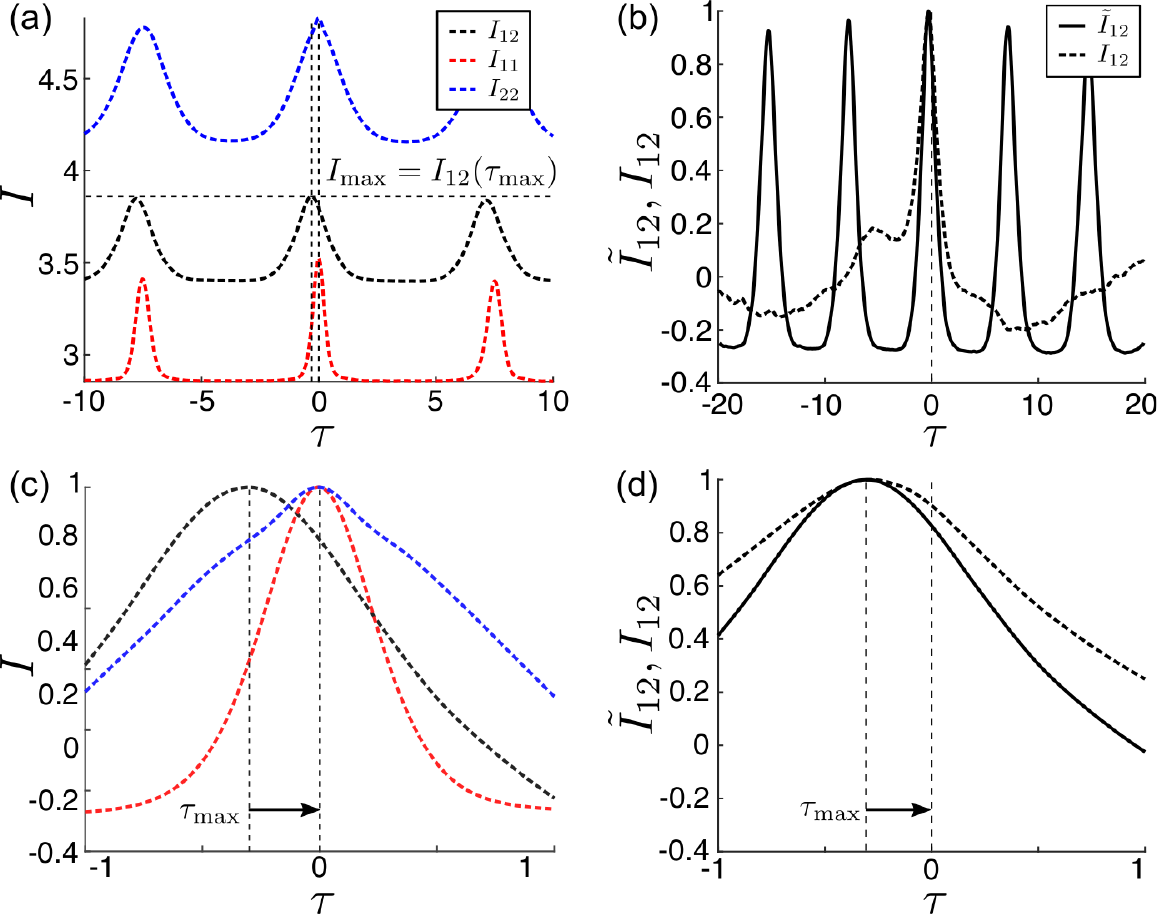}
    \caption{
    \label{fig:MutualInformation}
    Panel (a): Mutual informations $I_{\mu\nu}$ as a function of $\tau$ between and within the two subpopulations $\mu$ and $\nu$. Panel (c): The range-zoomed curve reveals a directionality in the mutual information $I_{12}(\tau)$ between the two subpopulations 1 and 2, with a maximum peak $\Imax \coloneqq I_{12}(\taumax)$ located at $\taumax < 0$. 
    Panels (b) and (d): comparison of the average mutual information obtained from continuous time data (solid), $\tilde{I}_{12}$, versus event time data (dotted), $I_{12}$. Obviously, locations of the peaks agree very well. 
    Panels (b-d) display the average mutual information $(I-\avt{I})/\max_t{(I-\avt{I})}$ with $\avt{I}$ denoting the average over time.  
    Parameters across panels are $A=0.275$, $\beta=0.125$, $\sigma_\omega = 0.01$, $\sigma_C = 0.01$, $\sigma_W^2 = 0.001$. 
    }
  \end{center}
\end{figure}

The first results for the event data outlined in Section \ref{sec:methods:poincare} are shown in Fig.~\ref{fig:MutualInformation}(a). Since the recurrence of events mimicked the mean frequency of the phase oscillators, the mutual information turned out periodic. As expected, we observed  (average) values of the mutual information that differed between $ I_{11}$ (red), $I_{22}$ (blue), $ I_{12} $ (black). This relates to the difference in entropy of the subpopulations, with the less synchronized one ($\mu=2$) being more disordered. The latter, hence, contained more entropy. However, since we were not interested in the explicit values of $I_{\mu\nu}(\tau)$, we could rescale the mutual information to maximum value which allowed for a comparative view when zooming in to the neighborhood of $\tau\approx 0 $; see panel (c) of Fig.~\ref{fig:MutualInformation}. The off-zero peak of $\Imax  \coloneqq  I_{12}(\taumax)$ in $\tau = \taumax $ clearly indicated a directed information flow, i.e. there is a \emph{functional directionality despite the structural symmetry} of our model.

When comparing the estimates for the mutual information obtained from the event data with those from the continuous-time we found little to no difference in peak location. That is,  the positions  $\taumax$ of the maximum peaks, $\Imax$ and $\tilde{I}_{\rm max}$, were nearly identical using either method as shown in panels (b) and (d) Fig.~\ref{fig:MutualInformation}. Thus, to study the effects of varying model parameters on $I_{\mu\nu}(\tau)$, our subsequent analysis may solely rely on the event data approach.

\begin{figure}[!htp]
\begin{center}
\includegraphics[width = \columnwidth]{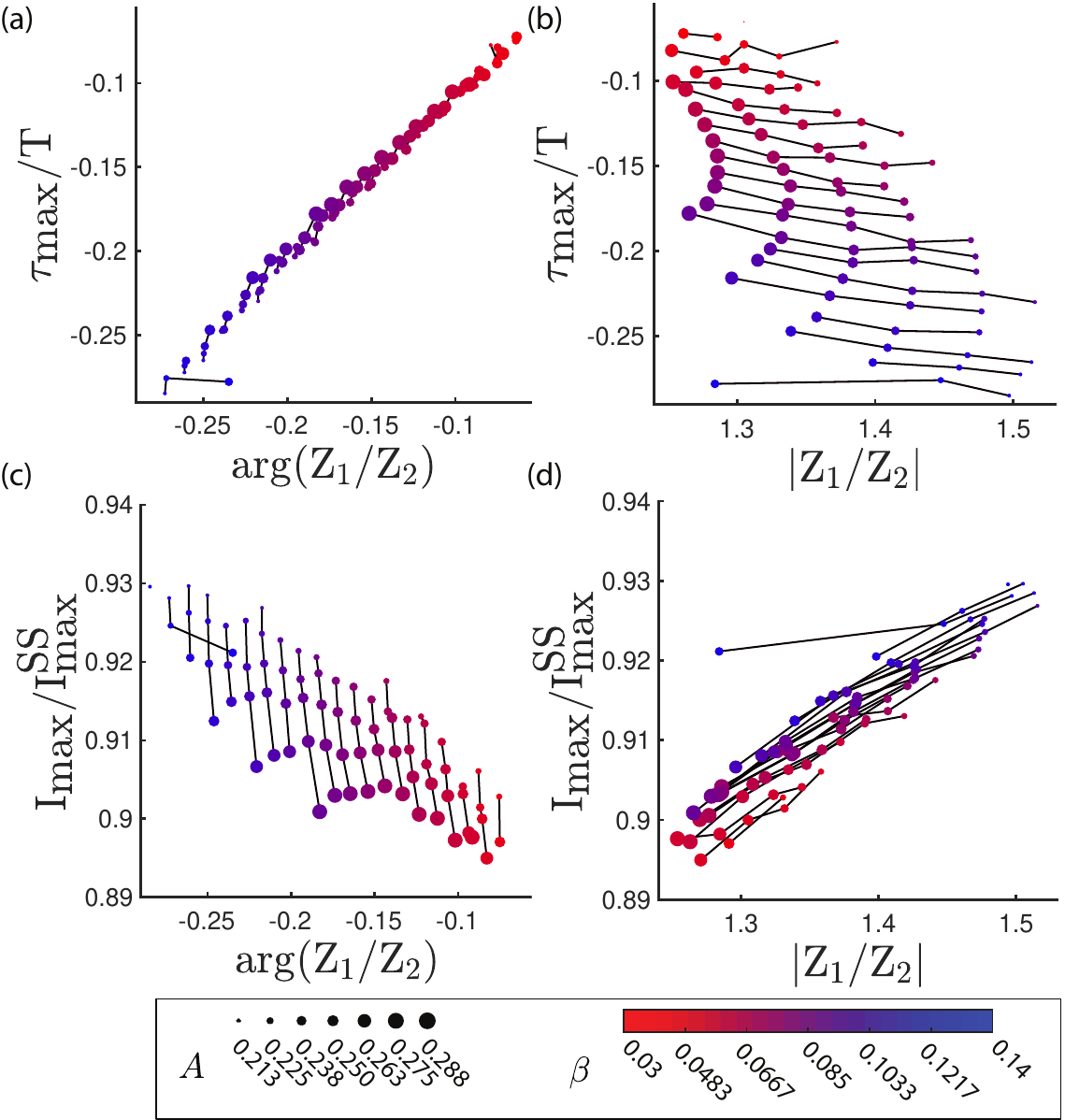}
\caption{
\label{fig:dependency}
Dependency of $(I_{\rm max},\nu_{\rm max})$ on the synchronization level and on the parameters $\beta$ and $A$.
Panel (a): The maximizing time delay $\taumax$ has a nearly linear dependency on the angular phase difference of order parameters $\arg{(Z_1/Z_2)}$.
Panel (b): $\taumax$ appears largely independent of the ratio of the magnitude of order parameters $|Z_1/Z_2|$, though
results are weakly dependent on $\beta$.
Panels (c) and (d): The normalized maximum peak of mutual information $\Imax/\ISSmax$ shows a weak dependency on the ratio of order parameters.
All other parameters are fixed at $\sigma_\omega = 0.01, \sigma_C = 0.01,\sigma_W^2 = 0.001$.
}
\end{center}
\end{figure}

Varying the values of $A$ and $\beta$ revealed a strong relation between the location of the peak of the delayed mutual information $\taumax$ and the relative phase $\arg{(Z_1/Z_2)}$ between the two subpopulations; see Fig.~\ref{fig:dependency}(a). This convincingly shows that our approach to analyze the event-based data reveals important information about the (otherwise continuous) dynamics of the subpopulations, here given by the phases (arguments) of the local order parameters. By contrast, the relative strength of local synchronization $|Z_1/Z_2|$ had little to no effect on $\taumax$; see Fig.~\ref{fig:dependency}(b). 
These  dependencies were inverted when looking at the value of mutual information, $\Imax/\ISSmax$. Consequently, the value of mutual information was affected by relative strength of local synchronization $|Z_1/Z_2|$ | after all the more a subpopulation is synchronized, the lower the corresponding entropy. However, effects were small and probably negligible when transferring to (selected) experimental data. More details about the normalization factor $\ISSmax$ are discussed  in \emph{Appendix}~\ref{app:normalization}.

\section{Discussion}
The delayed mutual information from event data, here, the passing times through the oscillators' Poincar\'e sections, agreed with that of the continuous data. This is an important finding as it shows that studying discrete event time series allows for inferring information-related properties of continuous dynamics. This offers the possibility to bring our approach to experimental studies, be that in neuroscience, where spike trains are a traditional measure of continuous neuronal activity, or economics, where stroboscopic assessments of stocks are common practice. Here, we used this approach to explore the information flow within and between subpopulations of a network of coupled phase oscillators. This information flow turned out to be directed.

Mutual information is the traditional measure of shared information between the two systems~\cite{Shannon1949}. Our findings relied on the introduction of a time delay $\tau$, which readily allows for identifying a direction of flow of information. This concept is much related to the widely used transfer entropy~\cite{schreiber2000measuring,wibral2013measuring}. In fact, transfer entropy is the delayed \emph{conditional} mutual information~\cite{dobrushin1959general,wyner1978definition}. It therefore differs from our approach primarily by a normalization factor. Since here we were not interested in the absolute amount of information (flow), we could simplify assessments and focus on delayed mutual information | adding the conditional part jeopardizes the approximation of probabilities, i.e., estimating transfer entropy in a reliable manner typically requires more data than our delayed mutual information. A detailed discussion of these differences is beyond the scope of the current study.

For our model we found that the delayed mutual information between the synchronized subpopulation and the less synchronized one peaked at a finite delay $\taumax \ne 0$. Hence, there was a directed flow of information between the two subpopulations. Since we found that  $\taumax < 0$ for $I_{12}(\tau)$, the direction of this flow was from the less synchronized subpopulation to the fully synchronized one, i.e. $2 \rightarrow 1$ or  {{\rm{D}}} $\rightarrow$ {{\rm{S}}}. In fact, the delay largely resembled the relative phase between the corresponding local order parameters, as shown in panel (a) of Fig.~\ref{fig:dependency}. We could not only readily identify the relative phase by encountering event data only, but our approach also allowed for attaching a meaningful interpretation in an information theoretic sense. This is promising since event data are, as stated repeatedly, conventional outcome parameters in many experimental settings. This is particularly true for studies in neuroscience, for which the quest on information processing is often central. There, networks are typically complex and modular. The complex collective dynamics switches at multiple scales, rendering neuronal networks especially exciting when it comes to information routing \cite{kirst2016dynamic}. As of yet, our approach does not allow for unraveling dynamics information routing. This will require extending the (time-lagged) mutual information to a time-dependent version, e.g., by windowing the data under study. We plan to incorporate this in future studies.

\section{Conclusion}
Estimating the delayed mutual information based on time points at which the individual phases passed through their respective Poincar\'e sections allows for identifying the information flow between subpopulations of networks. If the network displays chimera states, the information flow turns out to be directed. In our model of coupled phase oscillators, the flow of information was directed from the less synchronized subpopulation to the fully synchronized one since the first preceded the latter. Our approach is a first step to study information transfer between spike trains. It can be readily adopted to static well-defined modular networks and needs to be upgraded to a time dependent version to be applied to real, biological data.

\section*{Funding\label{sec:funding}}
This study received funding from the European Union's Horizon 2020 research and innovation program under the Marie Sk\l{}odowska-Curie grant agreement \#642563 (\href{cosmos-itn.eu}{COSMOS}).

\section*{Author contributions\label{sec:contributions}}
ND conducted the simulations and analysis and wrote the manuscript; AD  set up the study, incorporated the kernel density estimation and wrote the manuscript, DB contributed to the first concepts of the study and wrote the manuscript; EAM set up the study, conducted simulations and wrote the manuscript.

\section*{Acknowledgements\label{sec:acknowledgements}}
ND and AD want to thank to Rok Cestnik and Bastian Pietras for fruitful discussions.

\include{Deschle_etal_InformationInChimeras_Revision.bbl}

\appendix
\section{Stability diagram for chimera states\label{app:stabilitydiagram}}
\begin{figure}[!h]
\vspace{15 pt}
  \begin{center}
    \includegraphics[width = \columnwidth]{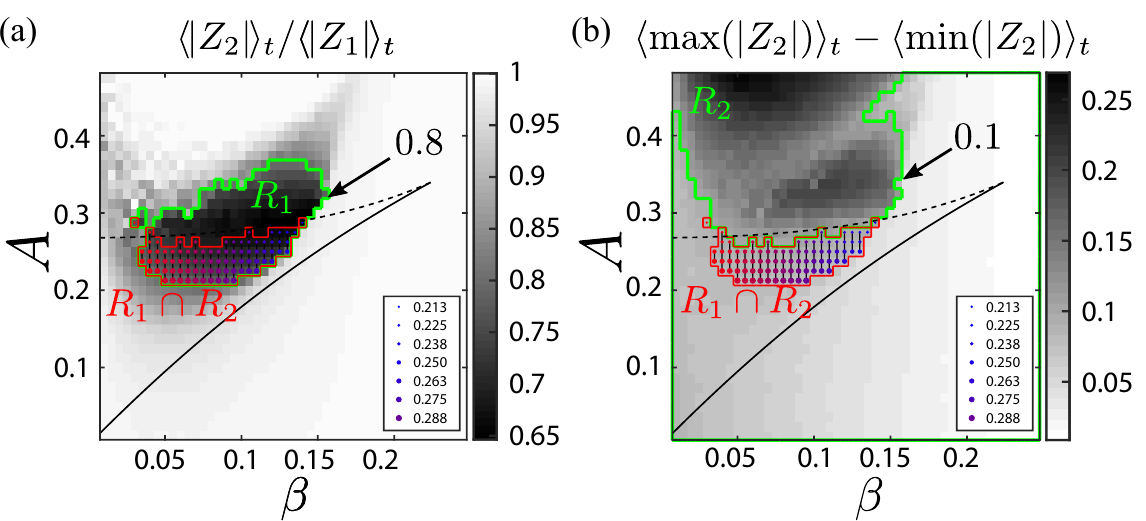}
    \caption{
    \label{fig:App1}
    Stability diagram based on analyzing complex-valued order parameters.
    Regions $R_1$ (a) and $R_2$ (b) in green outline parameter values $(\beta,A)$ for which $\avt{| Z_2 |} / \avt {|Z_1|} < 0.8$ and $\avt{\max(|Z_2|)}- \avt{ \min(|Z_2|) } < 0.1$, respectively. The states that are most similar to 'stable chimera states'~\cite{Abrams2008} were identified within the intersection $R_1\cap R_2$ (outlined in red). The outlined regions are bounded by parameter regions known to exhibit stable chimeras in the limit of $\sigma_\omega = \sigma_C = \sigma_W^2 = 0$~\cite{Abrams2008}; saddle node and Hopf bifurcation curves are shown as black solid and dashed curves, respectively.
    The other parameters have been fixed at $\sigma_\omega = 0.01$, $\sigma_C = 0.01$, and $\sigma_W^2 = 0.001$. 
    }
  \end{center}
\end{figure}

For the study of mutual information, we restricted ourselves to dynamic states most similar to the 'stable chimera states' reported in~\cite{Abrams2008}. 
Due to the various noise sources inherent to the model, the system dynamics undergoes fluctuations, as described in Section~\ref{sec:model}. This poses additional challenges that we addressed by examining specific time averages of the complex-valued order parameter in each subpopulation. 
Since fluctuations may cause the system to temporarily drift off the synchronized manifold, we may not simply consider $|Z_1|=1, |Z_2|<1$ to identify chimera states. Instead we require the order to be sufficiently different between the two subpopulations, i.e., we threshold $\avt{|Z_2|}/\avt{|Z_1|} $, defining the region $R_1$ shown in Fig.~\ref{fig:App1}(a). 

Furthermore, we presume amplitude variations to be small. To that end, we thresholded $\avt{\max(|Z_2|)} - \avt{\min(|Z_2|)}$, defining the region $R_2$ shown in Fig.~\ref{fig:App1}(b). Here, $\avt{\cdot}$ denotes the average defined over the maxima computed over time windows of length $ T=10^3 $ time steps, over a duration $T=9 \cdot 10^5$ after removing a transient of $T=10^5$ time steps.
Finally, chimera states with parameter values $(\beta,A)\in R_1\cap R_2$ were selected for further analysis of mutual information.

\section{Normalization factor $I_{\rm max}^\SS$ \label{app:normalization}}
\begin{figure}[!ht]
  \begin{center}
    \includegraphics[width = 0.5\columnwidth]{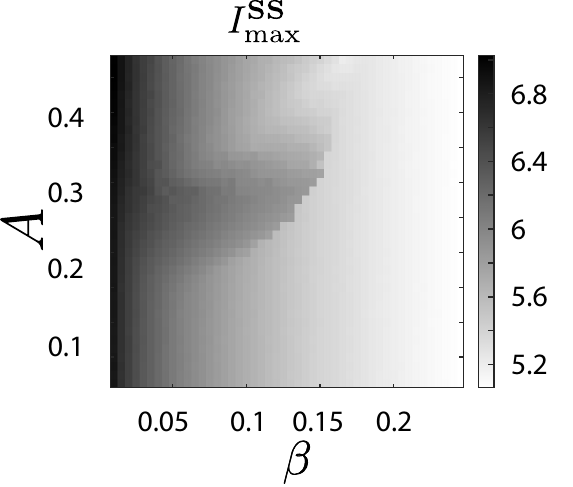}
    \caption{
    \label{fig:App2}
    The normalization factor given by the peak value of mutual information in a fully synchronized state $ \SS $, $ \ISSmax  \coloneqq  I_{11}(\tau_{\rm \max}) = I_{22}(\tau_{\rm \max} $) as  a function of $ (\beta,A) $.
    The other parameters are kept fixed at $\sigma_\omega = 0.01$, $\sigma_C = 0.01$, and $\sigma_W^2 = 0.001$.
  }
  \end{center}
\end{figure}

Our model displays a baseline dependency of the maximum peak of mutual information $I_{\rm max}^\SS$ on the parameters $\beta$ and $A$, even when the system is in the fully synchronized state $\SS$, see Fig.~\ref{fig:App2}.
We therefore studied the dependency of the maximum peak of the delayed mutual information for chimera states normalized with $\ISSmax$, $\Imax/\ISSmax$. For every set of parameters we computed the delayed mutual information on two different solution that have been forced by changing the initial conditions; the maxima of each of these solutions ($\SS$ and $\SD$) are $ \ISSmax \coloneqq  \ISSmax(A, \beta)$ and $ \Imax \coloneqq  \ISDmax(A,\beta) $. These are the quantities displayed in panels (c) and (d) of Fig.~\ref{fig:dependency}.

\end{document}

%% file: Deschle_etal_InformationInChimeras_Revision.bbl
%